\documentclass[12pt]{article}
\usepackage{pic02}
\usepackage{hyperref}
\usepackage{url}
\usepackage{graphicx}

\newcommand{\HZqqqq}{$\mathrm{HZ\rightarrow q\bar{q}q\bar{q}}$\hspace{1mm}}
\newcommand{\HZqqnn}{$\mathrm{HZ\rightarrow q\bar{q}\nu\bar{\nu}}$\hspace{1mm}}
\newcommand{\HZqqll}{$\mathrm{HZ\rightarrow q\bar{q}\ell^+\ell^-
(\ell = e,\mu)}$\hspace{1mm}}
\newcommand{\HZqqttqq}{$\mathrm{HZ\rightarrow q\bar{q}\tau^+\tau^-,
\tau^+\tau^-q\bar{q}}$\hspace{1mm}}
\newcommand{\qqg}{$\mathrm{e^+e^-\rightarrow q\bar{q}(\gamma)}$}
\newcommand{\WW}{$\mathrm{e^+e^-\rightarrow W^+W^-}$}
\newcommand{\ZZ}{$\mathrm{e^+e^-\rightarrow ZZ}$}
\newcommand{\Wen}{$\mathrm{e^+e^-\rightarrow We\nu}$}
\newcommand{\Zee}{$\mathrm{e^+e^-\rightarrow Ze^+e^-}$}
\newcommand{\Hbb}{$\mathrm{H\rightarrow b\bar{b}}$\hspace{1mm}}
\newcommand{\ra}{\rightarrow}
\newcommand{\hZMSSM}{$\mathrm{e^+e^-\ra hZ}$\hspace{2mm}}
\newcommand{\etal}{{\it et al.}}
\begin{document}

\title{\bf HIGGS SEARCH RESULTS}
\author{
Alexei Raspereza \\
{\em DESY, Platanenallee 6, D-15738 Zeuthen, Germany}}
\maketitle


%
%
%
\vspace{4.5cm}
%

\baselineskip=14.5pt
\begin{abstract}
This paper shortly reports on the results of Higgs boson searches 
performed at LEP and Tevatron. No signal is found and 
limits on the mass and couplings of the Higgs boson are 
derived. Interpreting the data in the framework of the Standard Model,
the 95\% C.L. lower limit of 114.4 GeV on the Higgs boson mass 
is obtained.
\end{abstract}
\newpage

\baselineskip=17pt

\section{Introduction}
The Standard Model (SM)~\cite{sm} of electroweak interactions 
is a well established theory which has been very
successful in accounting for various experimental 
observations as well as in its predictions of new 
physics phenomena. The unproven keystone of the SM 
is the mechanism of mass generation. The most favoured 
one is the so-called Higgs mechanism~\cite{higgs_mech}.
A doublet of scalar complex 
fields with non-zero vacuum expectation value is postulated,
breaking electroweak symmetry and generating 
weak boson and fermion masses. 
The Higgs mechanism gives rise to one more 
physical state $-$ the Higgs boson. The discovery of this 
particle would be a crucial step in establishing the mechanism 
of electroweak symmetry breaking and in our understanding 
of the origin of mass. One Higgs doublet is the minimum which is required
to generate weak boson and fermion masses. There are
extensions of the SM postulating additional Higgs multiplets
and thus predicting a spectrum of physical Higgs
particles~\cite{higgs_hunter}. Over the last decade searches 
for Higgs bosons of various 
theoretical models have been performed at the Tevatron 
$\mathrm{p\bar{p}}$ collider and the Large Electron-Positron (LEP)
collider. The results of these searches are 
reviewed. 

\section{The Standard Model Higgs Boson}

\subsection{Search for the SM Higgs Boson at LEP}
Over the last three years
of LEP running, the four LEP collaborations, ALEPH, 
DELPHI, L3 and OPAL, have collected 2460 pb$^{-1}$
of data at centre-of-mass energies between 189 and 209 GeV 
with about 530 pb$^{-1}$ above 206 GeV. These data provided
sensitivity to the SM Higgs boson up to a mass 
of $\mathrm{m_H}$ $\sim$ 115 GeV. 

\begin{figure}[h]
\begin{center}
\begin{tabular}{cc}
\hspace{-4mm}
\includegraphics*[width=0.51\textwidth]{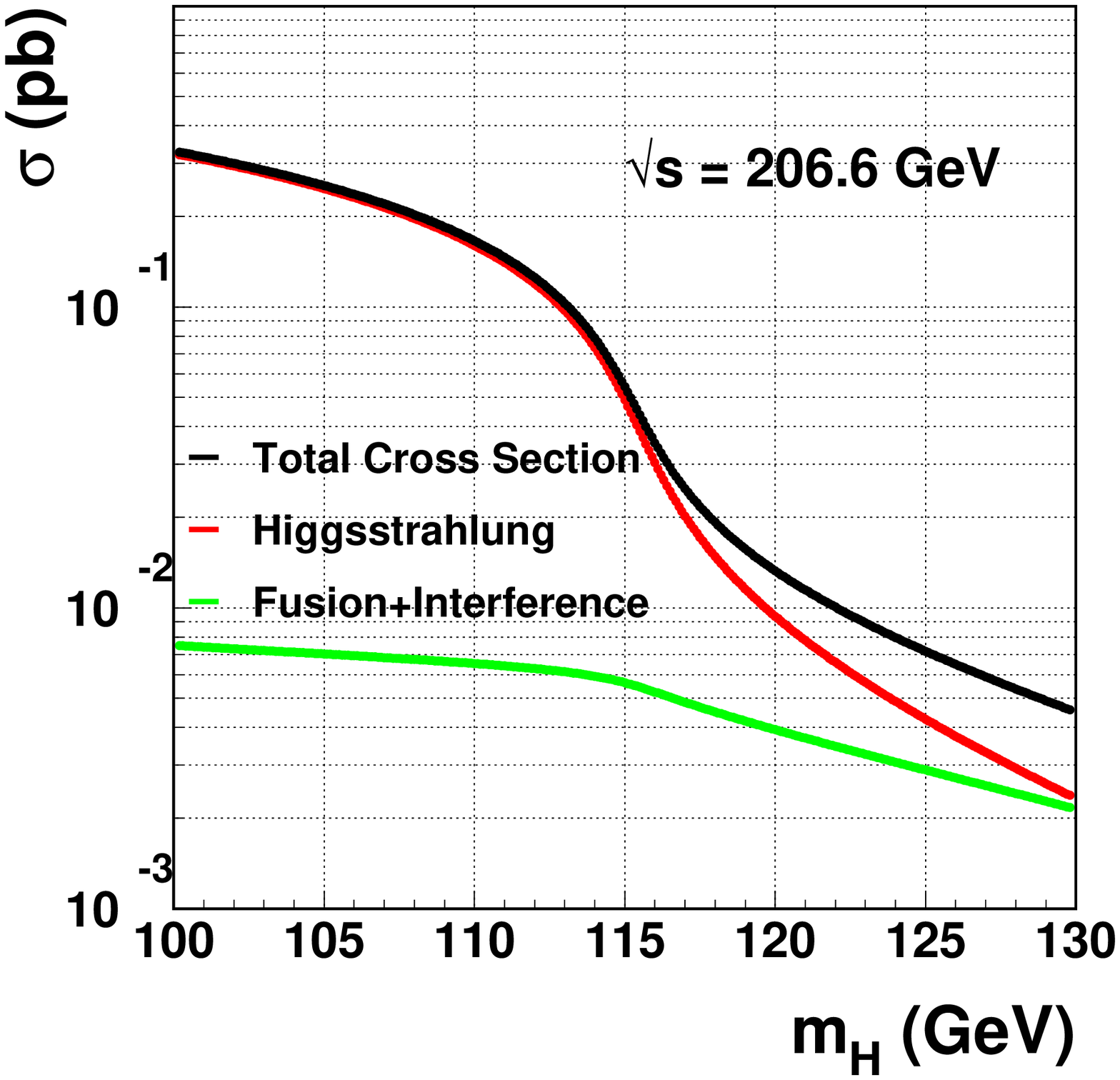} &
\hspace{-4mm}
\vspace{2mm} 
\includegraphics*[width=0.49\textwidth]{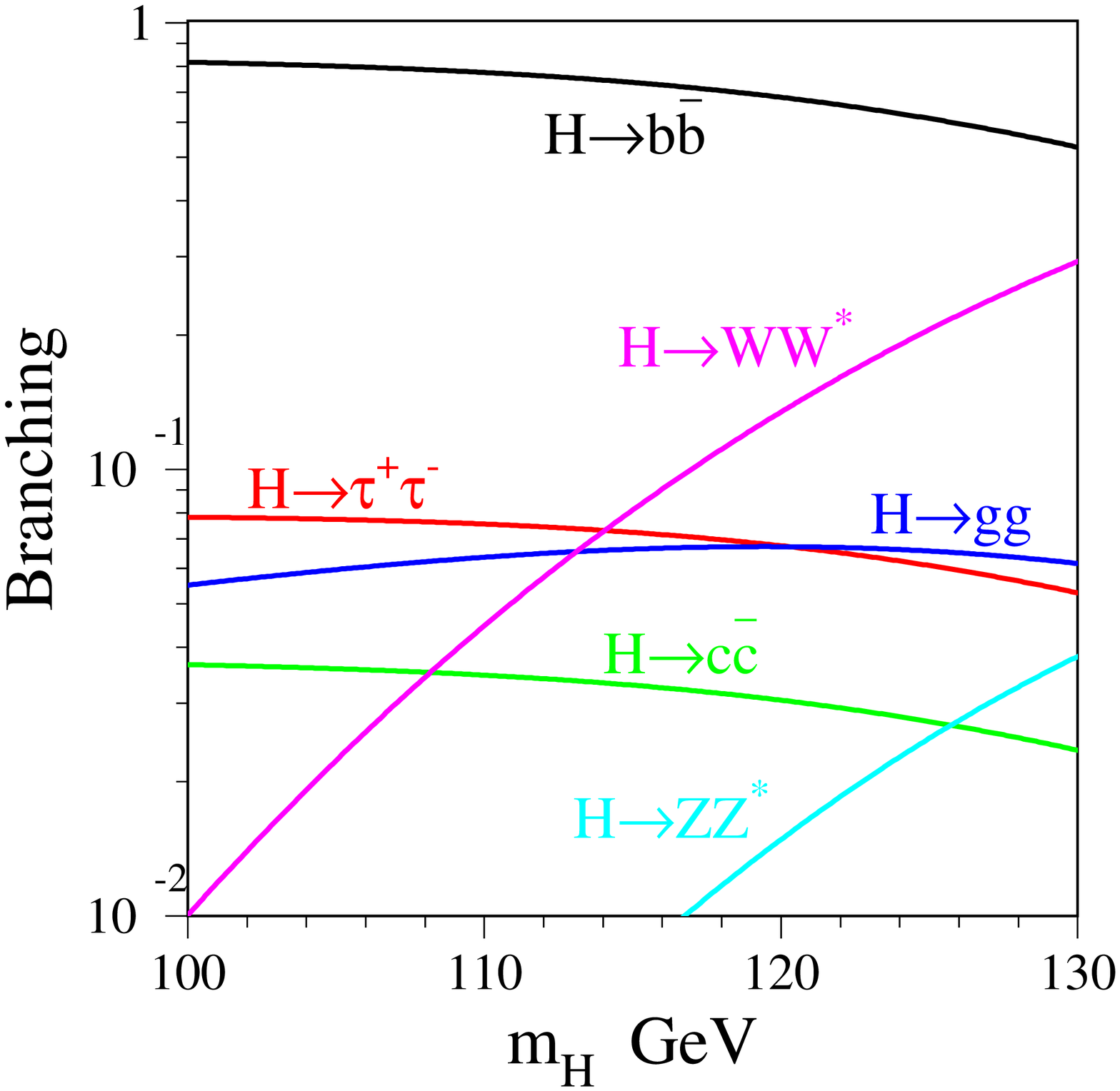} 
\end{tabular}
\end{center}
\caption{The production and decays of the SM Higgs boson.
Left: the cross sections of the production mechanisms at LEP
at $\mathrm{\sqrt{s}}$ = 206.6 GeV as a function of $\mathrm{m_H}$.
Right: the decay branching fractions as
a function of $\mathrm{m_H}$.}
\label{fig_sm_hzbran}
\end{figure}

At LEP, the production of the Higgs boson is expected 
mainly via the Higgs-strahlung process, 
$\mathrm{e^+e^-\ra Z^*\ra HZ}$. The processes of 
WW and ZZ fusion, 
giving rise to the $\mathrm{H\nu\bar{\nu}}$ and $\mathrm{He^+e^-}$
final states, contribute with a smaller rates
to the Higgs boson production. The dependence of the production cross 
sections on the Higgs 
boson mass at a typical LEP energy, attained in the year 2000, is 
illustrated in Figure~\ref{fig_sm_hzbran} (left). The SM Higgs boson 
decay branching fractions as a function of the Higgs boson mass 
are displayed in Figure~\ref{fig_sm_hzbran} (right).

The search for the SM Higgs boson is based on the study of four distinct
topologies, arising from the Higgs-strahlung process:
\HZqqqq (four-jet channel);
\HZqqnn (missing energy channel);
\HZqqll (semileptonic channel) and
\HZqqttqq (tau channels).
Searches in channels with hadronic decays of the 
Higgs boson are optimised for the \Hbb decay mode, which is 
predicted to be dominant in the mass range accessible at LEP.
Hence, tagging b quarks plays a crucial role in identification 
of signal events.
The main processes contributing to the background for the Higgs 
boson searches are:  
quark antiquak pair production with possible initial 
state radiation, \qqg, 
pair-wise production of W bosons, \WW,
pair-wise production of Z bosons, \ZZ,
single W boson production, \Wen, and
single Z boson production, \Zee.

\begin{figure}[h]
\begin{minipage}[c]{0.45\textwidth}
  \includegraphics[width=1.1\textwidth]{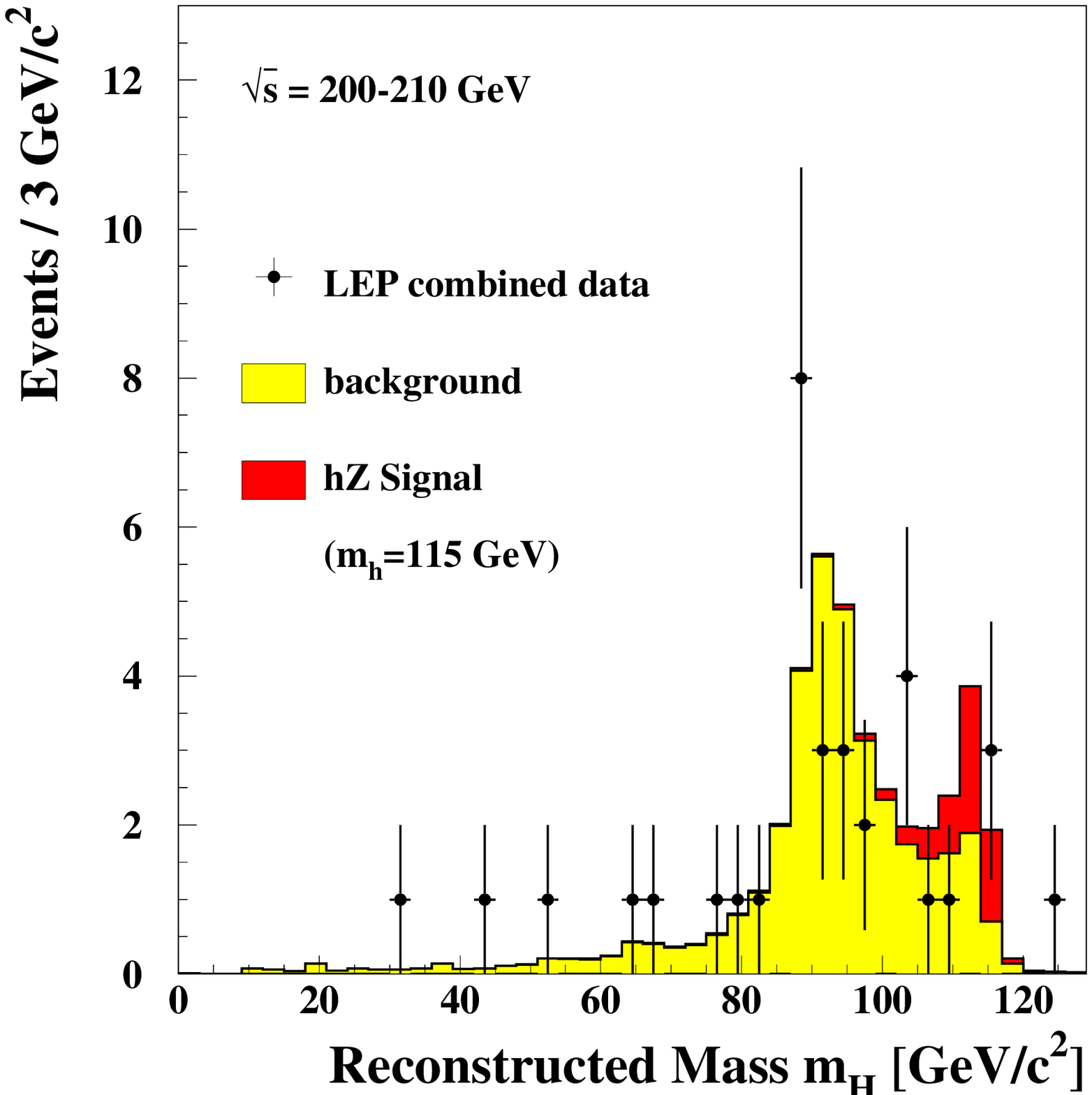}
    \caption{\label{fig:smmass}
            Distribution of the reconstructed Higgs boson 
            mass, $\mathrm{m_H^{rec}}$, obtained with 
            a special, non-biasing, selection. The cuts 
            are adjusted in such a way as to obtain 
            approximately equal numbers of expected 
            signal and expected background events 
            in the region $\mathrm{m_H^{rec}}$ $>$ 109 GeV.
            }
\end{minipage}
\hspace{4mm}
\begin{minipage}{0.05\textwidth}
\end{minipage}
\begin{minipage}[c]{0.5\textwidth}
\hspace{-5mm}\includegraphics[width=1.2\textwidth]{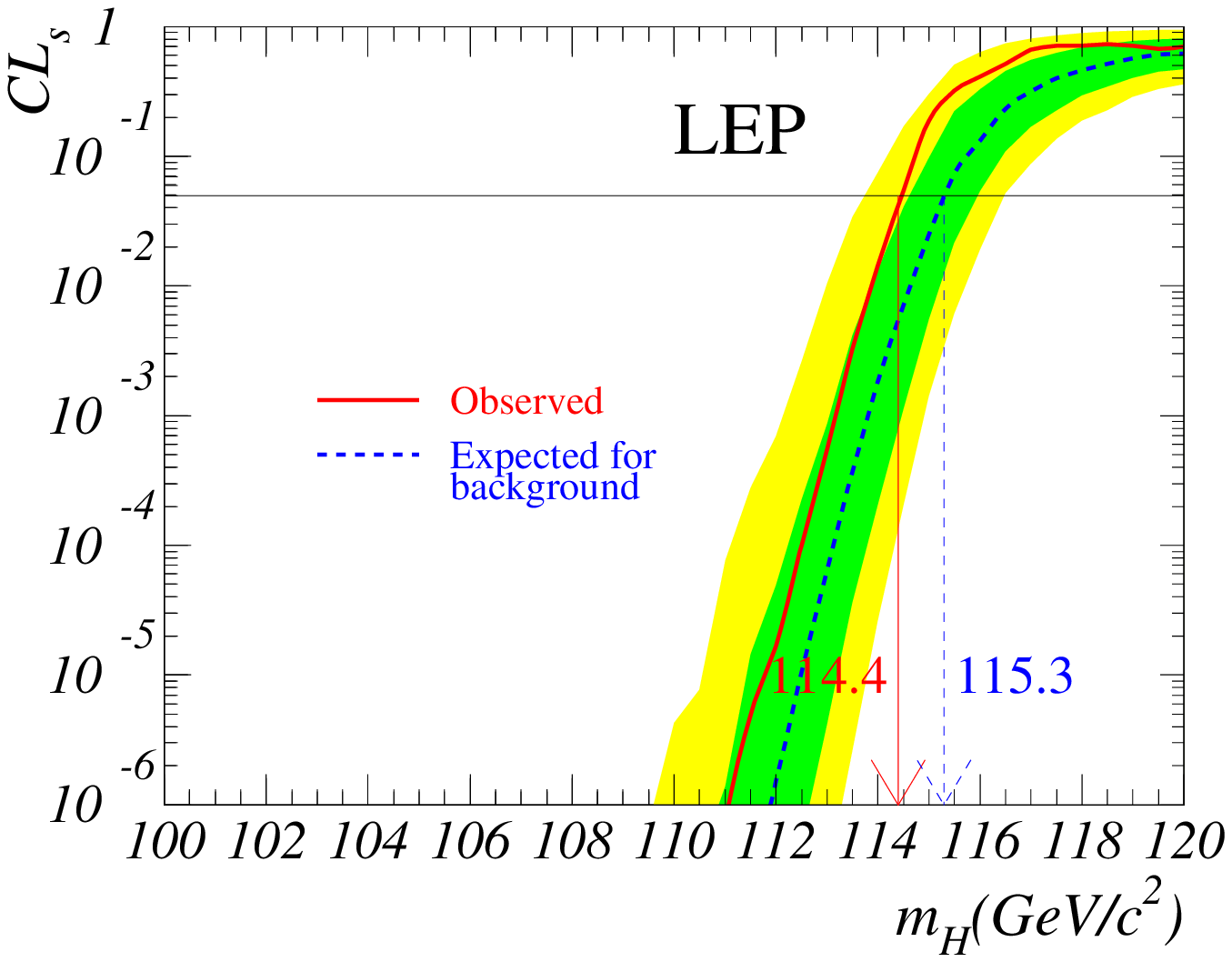}
\caption{\label{fig:cls}
        Confidence level $\mathrm{CL_s}$ as a function of the 
        tested Higgs boson mass. The solid line: observation;
        the dashed line: expectation in the absence of a signal.
        The shaded areas represent $1\sigma$ and $2\sigma$ bands 
        centred on the value expected in the absence of a signal.
        Results from the four LEP collaborations are combined.
        }
\end{minipage}
\end{figure}

The results of the individual analyses are expressed in terms
of one (or more) final variable, also called {\it{discriminant}}.
It is constructed, exploiting event features, which discriminate 
between background and signal. Among them are the invariant mass
of the two jets (or two tau leptons) assumed to stem from the 
Higgs boson, hereafter called
the reconstructed Higgs boson mass, $\mathrm{m_H^{rec}}$, 
b-tag variable, quantifying the probability for an
event to contain b jets, and topological characteristics of an event.
As an example, Figure~\ref{fig:smmass} shows the distribution 
of $\mathrm{m_H^{rec}}$ after combining data from all four 
experiments in all search channels.
The final variable, built in the same way for data, Monte Carlo background 
and Monte Carlo signal samples, is then used to evaluate on statistical 
basis the presence of a signal in data. 

An excess of $3\sigma$ beyond background expectation compatible 
with the production of the Higgs boson with the mass near 
115 GeV is reported by ALEPH collaboration~\cite{aleph_sm}. 
This effect comes 
from events of the \HZqqqq topology and is not confirmed neither
by other search channels nor by other LEP 
collaborations~\cite{dlo_sm}.
In the LEP combined search, the excess is diminished down 
to about $1.7\sigma$~\cite{sm_lep} due to 
background-like
observations by the other experiments.
The LEP combined data are used to set a lower 
limit on the Higgs boson mass. Figure~\ref{fig:cls} shows 
the dependence of the confidence level $\mathrm{CL_s}$, used 
to exclude the signal hypothesis, on the tested Higgs boson mass.
An observation of the $\mathrm{CL_s}$ value smaller than 5\% 
means that the signal hypothesis is ruled out at 95\% C.L.. 
A lower limit of 114.4 GeV on the mass
of the SM Higgs boson is derived at 95\% C.L..

\subsection{Searches for the SM Higgs Boson at Tevatron}
The searches for the SM Higgs boson are performed
by the CDF and D0 collaborations~\cite{sm_teva} using about 0.1 fb$^{-1}$ 
of data collected per experiment 
at centre-of-mass energy $\mathrm{\sqrt{s}}$ = 1.8 TeV during 
the RunI data taking period at Tevatron. 

\begin{center}
\begin{figure}[h]
\begin{minipage}[c]{.64\textwidth}
\mbox{\includegraphics[angle=270, width=0.99\textwidth]{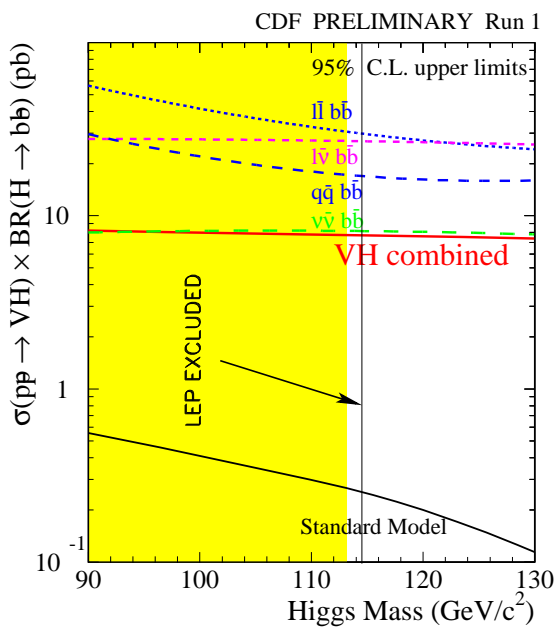}}
\end{minipage}
\begin{minipage}[c]{.34\textwidth}
\vspace{2.0cm}
\caption{The 95\% C.L. upper limit set by CDF on the product of
the HV (V is W and Z) production cross section and 
the branching fraction of $\mathrm{H\ra b\bar{b}}$ compared to 
the SM prediction. Also shown is the combined LEP limit. }
\label{fig:CDF_limits}
\end{minipage}
\end{figure}
\end{center}

Although single Higgs boson production via gluon fusion, 
$\mathrm{gg\rightarrow H}$, has the highest cross-section, 
the more promising channels for the SM Higgs boson search are 
$\mathrm{q\bar{q}^\prime\rightarrow HW}$ 
and $\mathrm{q\bar{q}\rightarrow HZ}$ production, since 
a large fraction of copious QCD background can be suppressed
by tagging leptons originating from decays of W or Z bosons. 
The Higgs boson is searched for by its decay
into $\mathrm{b\bar{b}}$.
Four final states are studied:
$\mathrm{HZ\rightarrow b\bar{b}\ell^+\ell^-(\ell=e,\mu)}$;
$\mathrm{HZ\rightarrow b\bar{b}\nu\bar{\nu}}$;
$\mathrm{HW\rightarrow b\bar{b}\ell\nu(\ell=e,\mu)}$ and
$\mathrm{HZ(HW)\rightarrow b\bar{b}q\bar{q}}$.
Data are found to agree with the SM background predictions.
However, the sensitivity of this search does not reach
the SM expectations for a signal: the 95\% C.L. 
upper limit on the quantity  
$\mathrm{\sigma(p\bar{p}\ra HV)\times Br(H\ra b\bar{b})}$ 
(where V stands for W and Z) is larger than 
the values predicted by the SM, as can be seen in Figure~\ref{fig:CDF_limits}. 
Hence, a mass limit cannot be derived.

\section{Searches for Higgs Bosons beyond the SM}
Among all possible extensions of the SM, Two Higgs 
Doublet Models (2HDM)~\cite{higgs_hunter} are of particular 
interest, since this structure of the Higgs sector is required 
in ``low-energy'' supersymmetric models. 
The Higgs sector of 2HDM is characterised by eight 
degrees of freedom. After spontaneous symmetry breaking, 
three of them become longitudinal polarisations of W and Z
bosons, while the remaining five manifest themselves as 
physical states: two neutral CP-even Higgs bosons, the lighter of 
which is denoted as h and the heavier as H,
one CP-odd Higgs boson, A, and a pair of charged bosons, $\mathrm{H^\pm}$.
At this point one should mention two parameters of 
2HDM, which play a crucial role in the Higgs boson phenomenology. These are
$\tan\beta$, the ratio of vacuum expectation values 
of the two Higgs doublets, and $\alpha$, the mixing angle
in the CP-even Higgs sector. 

\subsection{Flavor Independent Search for $\mathrm{e^+e^-\ra hZ}$}
\begin{center}
\begin{figure}[h]
\begin{minipage}[c]{.55\textwidth}
\mbox{\includegraphics[angle=0, width=0.98\textwidth]{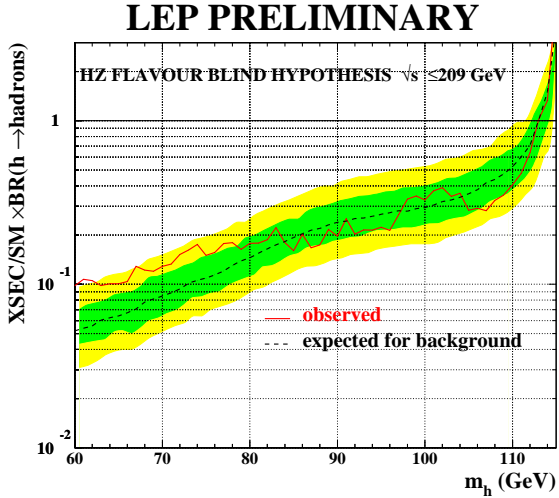}}
\end{minipage}
\begin{minipage}[c]{.45\textwidth}
\vspace{-1.0cm}
\caption{\label{fig:lep_flvb}The 95\% C.L. upper limit on 
         the Higgs-strahlung cross section, 
         normalised to the value expected in the SM, 
         as a function of the tested Higgs boson mass.
         The limit is obtained from the LEP combined 
         flavor independent search for the \hZMSSM process
         assuming $\mathrm{B(h\ra hadrons)}$ equal 1.
         The solid line represents the observed limit, the dashed line
         stands for the expected limit and the shaded areas show 
         $1\sigma$ and $2\sigma$ bands centred on the expected 
         limit .
}
\end{minipage}
\end{figure}
\end{center}

The distinct feature of 2HDM is the modified couplings of the Higgs bosons
to fermions and gauge bosons. 
Table~\ref{tab:2hdm} illustrates the dependence of the neutral 
Higgs boson couplings to fermions on the parameters $\alpha$ and $\beta$
in 2HDM of type II, in which one Higgs doublet couples 
to up-type quarks while the other one to down-type quarks and charged leptons.
\begin{table}[h]
\begin{center}
\begin{tabular}{|c|c|c|c|}
\hline
                        & h & H & A \\
\hline
 $\mathrm{u\bar{u}}$ (up-type quarks)  &   
 $\mathrm{\cos\alpha/\sin\beta}$ &   
 $\mathrm{\sin\alpha/\sin\beta}$ &   
 $\mathrm{\cot\beta}$ \\
\hline
 $\mathrm{d\bar{d}}$ (down-type quarks) &
 $\mathrm{\sin\alpha/\cos\beta}$ &   
 $\mathrm{\cos\alpha/\cos\beta}$  &   
 $\mathrm{\tan\beta}$ \\
\hline
 $\mathrm{\ell\bar{\ell}}$ (charged leptons) &
 $\mathrm{\sin\alpha/\cos\beta}$   &   
 $\mathrm{\cos\alpha/\cos\beta}$    &   
 $\mathrm{\tan\beta}$ \\
\hline  
\end{tabular}
\end{center}
\caption{\label{tab:2hdm}The dependence of
fermion couplings to the neutral Higgs bosons 
on the parameters $\alpha$ and $\beta$ in 2HDM 
of type II.}
\end{table}
As one can see, specific choices of parameters 
$\alpha$ and $\beta$ are possible, which lead to 
the suppression of the Higgs boson couplings to
$\mathrm{b\bar{b}}$. SM Higgs searches would have a reduced
sensitivity in such cases, because of their strong reliance 
on identification of b-quarks. To investigate these 
specific scenarios, all four LEP collaborations have
performed flavor independent searches for 
hadronic decays of the Higgs bosons~\cite{adlo_flindep}. 
The preliminary LEP combination of the search 
results is done for the Higgs-strahlung
production mechanism\footnote{ Flavor independent searches 
were also performed for the $\mathrm{e^+e^-\ra hA}$ production mechanism
individually by DELPHI, L3 and OPAL collaborations. 
However, no LEP combined results of these searches 
are available yet.}~\cite{lep_flindep}.
All decay channels of the Z boson
are studied and decays of the Higgs boson
into $\mathrm{b\bar{b}}$, $\mathrm{c\bar{c}}$ and a pair of gluons
are considered. None of the LEP collaborations observes an indication of 
a signal and the search results are translated into the 95\% C.L. 
upper limits on the Higgs-strahlung cross section.
This is shown in Figure~\ref{fig:lep_flvb}. 
Assuming SM cross section
for the Higgs-strahlung process and the hadronic branching fraction 
of the Higgs boson equal 1, a lower bound of 112.9 GeV on the 
Higgs boson mass is obtained at 95\% C.L. regardless
of the flavor content of the Higgs boson decay products. 

\subsection{Searches for Charged Higgs Bosons}
\begin{figure}[h]
\begin{minipage}[c]{0.56\textwidth}
\includegraphics[width=0.99\textwidth]{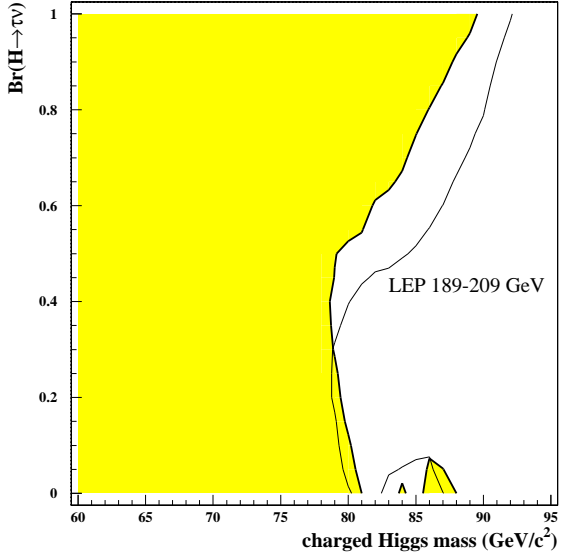}
\end{minipage}
\begin{minipage}[c]{0.42\textwidth}
\caption{\label{fig:hpm_lep}
The excluded region in the 
$\mathrm{(B(H^+\ra \tau^+\nu),m_{H^+})}$ plane.
Preliminary results from the four LEP collaboration are 
combined. The shaded area is experimentally excluded
at 95\% C.L.. The solid line indicates the expected 
boundary of the excluded region in the absence of a signal.}
\end{minipage}
\end{figure}

At LEP, charged Higgs bosons are expected to be produced 
in pairs through virtual Z or $\gamma$ exchange: 
$\mathrm{e^+e^-\ra Z^*/\gamma^* \ra H^+H^-}$. 
In 2HDM, 
the $\mathrm{H^\pm}$ mass is not predicted and the tree-level cross-section
is fully determined by the mass. The searches are carried out 
under the assumption that the two decay modes $\mathrm{H^+\ra c\bar{s}}$
and $\mathrm{H^+\ra \tau^+\nu}$ exhaust the $\mathrm{H^+}$ decay width
but the relative branching ratio is regarded as a free parameter.
Thus, the searches encompass the following final states:
$\mathrm{c\bar{s}\bar{c}s}$, $\mathrm{\tau^+\nu\tau^-\bar{\nu}}$
and $\mathrm{c\bar{s}\tau^-\bar{\nu}+\bar{c}s\tau^+\nu}$. 
An excess of candidates in the pure hadronic channel in the mass
region around 68 GeV is found in the L3 data~\cite{l3_hpm}. 
However, this effect 
is not confirmed by other LEP experiments~\cite{ado_hpm}.
No signal is found in the pure leptonic 
and mixed channels. The results from the four collaborations
are combined and used to set the 95\% C.L. lower limit
on the $\mathrm{H^+}$ mass in dependence of the branching fraction
$\mathrm{B(H^+\ra\tau^+\nu)}$, as shown in Figure~\ref{fig:hpm_lep}.
A lower limit of 78.6 GeV on the $\mathrm{H^+}$ mass is obtained
at 95\% C.L. independent of $\mathrm{B(H^+\ra \tau^+\nu)}$.

\begin{figure}[h]
\begin{minipage}[c]{0.45\textwidth}
\includegraphics[width=1.0\textwidth]{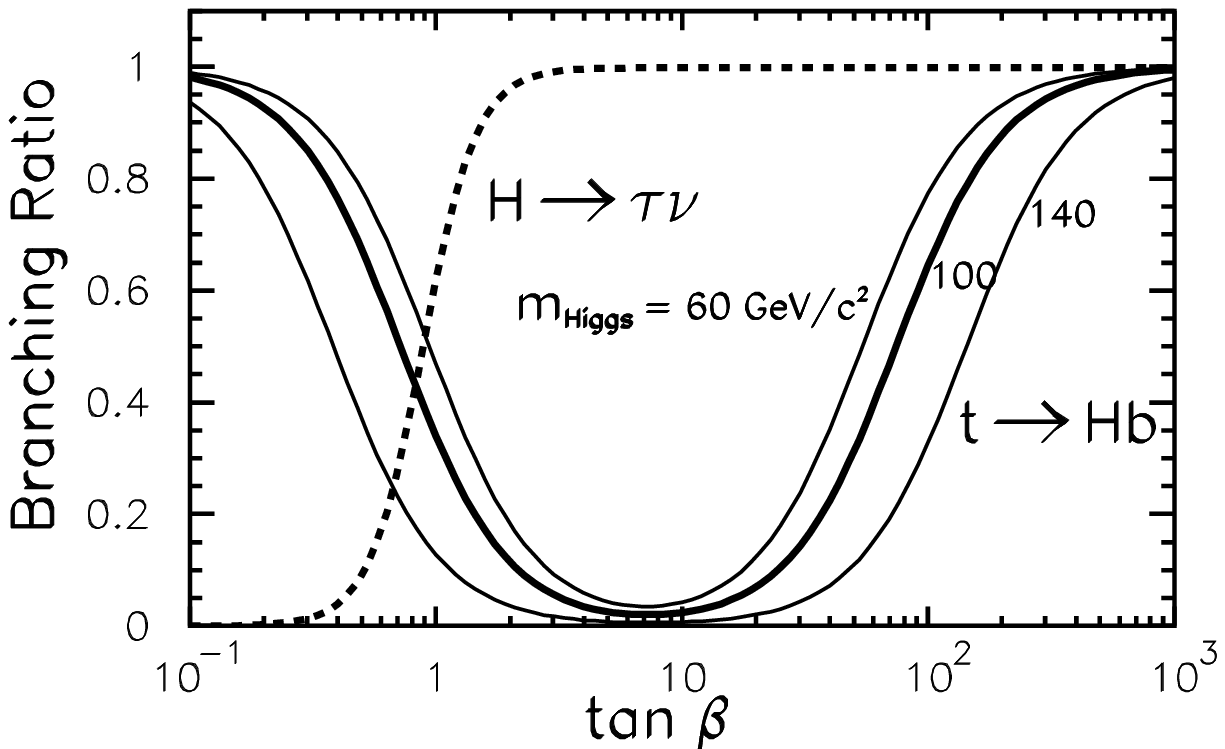}
    \caption{\label{fig:br_hpm}            
            Branching fractions 
            $\mathrm{B(t\ra H^+b)}$ and 
            $\mathrm{B(H^+\ra \tau^+\nu)}$
            as a function of $\tan\beta$.
            }
\end{minipage}
\hspace{4mm}
\begin{minipage}{0.05\textwidth}
\end{minipage}
\begin{minipage}[c]{0.50\textwidth}
\vspace{2mm}
\includegraphics[width=1.0\textwidth]{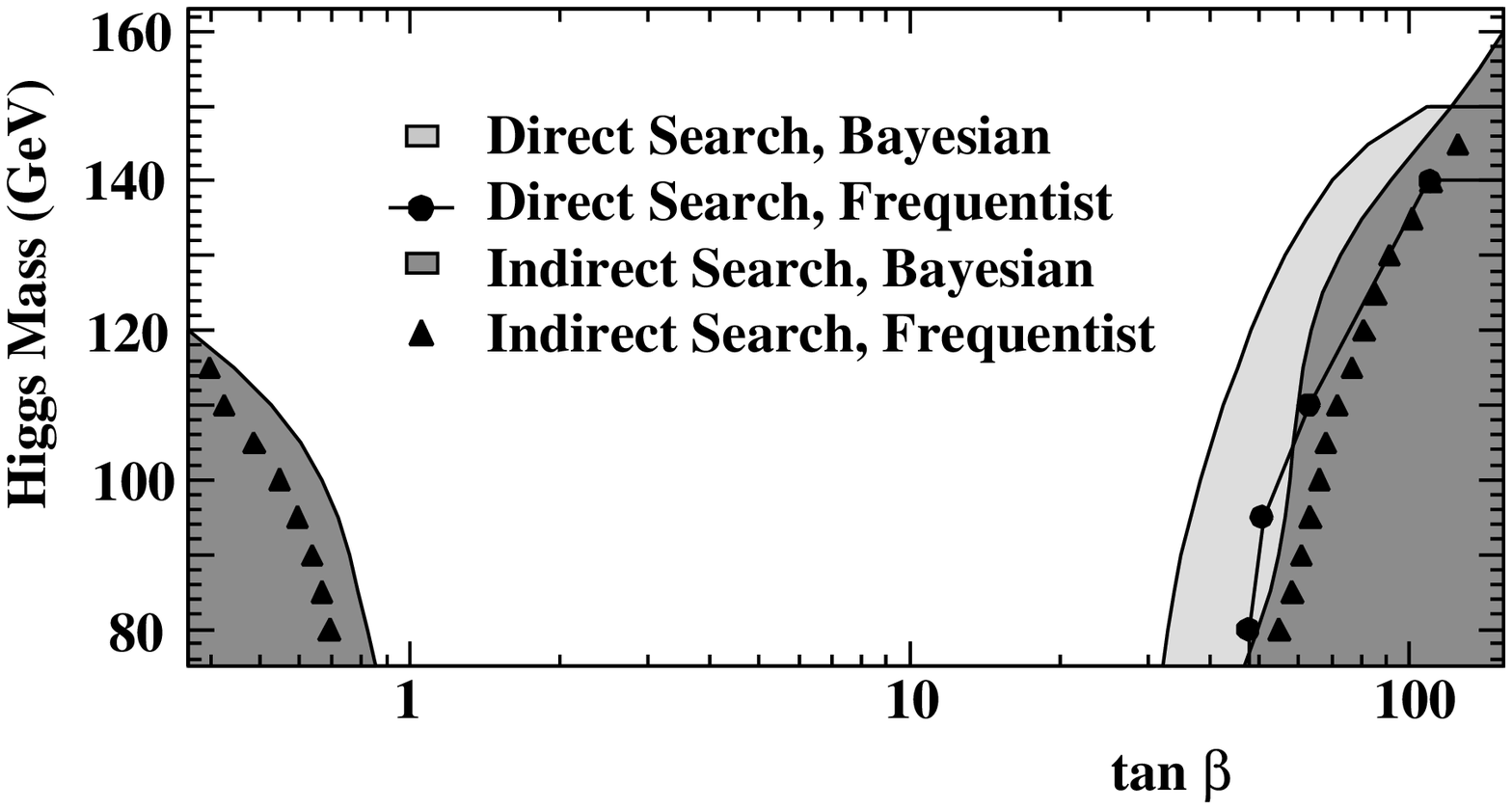}
\caption{\label{fig:d0_hpm}
        The 95\% C.L. excluded regions in the 
        $\mathrm{(m_{H^+},\tan\beta)}$ plane for the top quark mass 
        $\mathrm{m_t}$ = 175 GeV and 
        $\mathrm{\sigma(p\bar{p}\ra t\bar{t})}$ = 5.5 pb.
        Results are obtained by the D0 collaboration.
        }
\end{minipage}
\end{figure}

In the SM, the primary decay of the top quark is 
$\mathrm{t\ra W^+b}$. The addition of the second Higgs 
doublet allows the $\mathrm{t\ra H^+b}$ mode, provided
that $\mathrm{m_t>m_{H^+}+m_b}$. The dependence
of the branching fraction $\mathrm{B(t\ra H^+b)}$ on $\tan\beta$ is presented
in Figure~\ref{fig:br_hpm} for several representative values of 
the charged Higgs boson mass. $\mathrm{B(t\ra H^+b)}$ has a minimum
at $\mathrm{\tan\beta = \sqrt{m_t/m_b}}$. If $\mathrm{\tan\beta}$ 
differs by about a factor of 10 from $\mathrm{\sqrt{m_t/m_b}}$, 
the branching fraction becomes large, and decreases as $\mathrm{m_{H^+}}$
increases. Two different searches for $\mathrm{t\ra H^+b}$ are 
performed at Tevatron. An indirect search~\cite{hpm_teva_indirect} 
makes use of the fact that a large branching fraction $\mathrm{B(t\ra H^+b)}$ 
suppresses decay rates of $\mathrm{t\bar{t}}$
into the $\mathrm{W^+W^-b\bar{b}}$ final states. 
Hence, an observation of a decrease in the 
$\mathrm{t\bar{t}\ra W^+W^-b\bar{b}}$
signal expected from the SM would be an indirect indication 
of the $\mathrm{t\ra H^+b}$ mode. Direct searches~\cite{hpm_teva_direct} aim 
to select the $\mathrm{t\bar{t}\ra H^+bH^-\bar{b}}$ and 
$\mathrm{t\bar{t}\ra H^\pm W^\mp b\bar{b}}$ final 
states, exploiting the 
$\mathrm{H^+\ra \tau^+\nu}$ decay mode. The latter dominates 
at high $\tan\beta$ values, as shown in Figure~\ref{fig:br_hpm}. 
The decay mode $\mathrm{H^+\ra c\bar{s}}$ is not useful 
for direct searches due to a large QCD background. Dedicated analyses, 
performed at Tevatron, assume that there are no top quark decays 
other than 
$\mathrm{t\ra W^+b}$ and $\mathrm{t\ra H^+b}$. No signal is found
and the results are
used to constrain parameters of 2HDM. As an example, 
Figure~\ref{fig:d0_hpm}
presents D0 results, expressed in terms of the 95\% C.L. excluded regions 
in the $\mathrm{(m_{H^+},\tan\beta)}$ plane.

\subsection{Searches for the Neutral Higgs Bosons of the MSSM}
Supersymmetry is a tempting theoretical concept 
which provides an elegant solution of the so-called 
``hierarchy'' problem and incorporates 
gravity in a consistent way into the model~\cite{susy}. 
The Higgs sector of the 
Minimal Supersymmetric extension of the Standard Model
(MSSM)~\cite{mssm} corresponds to the 2HDM of type II and hence
contains the five physical states indicated above.

\begin{figure}[h]
\begin{minipage}[c]{.56\textwidth}
\includegraphics[angle=0, width=0.98\textwidth]{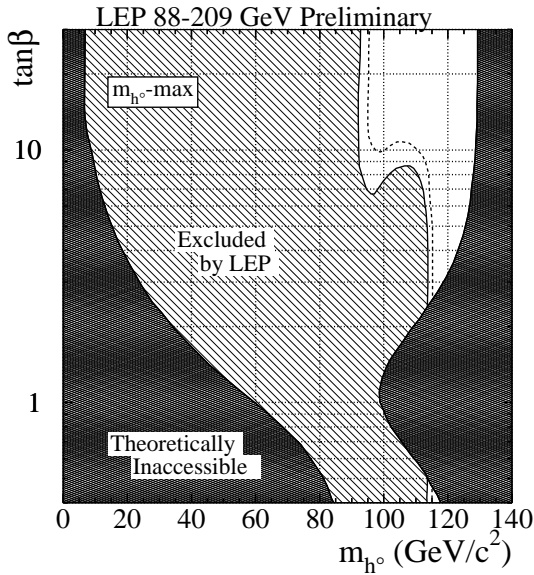}
\end{minipage}
\begin{minipage}[c]{.42\textwidth}
\vspace{-1.0cm}
\caption{\label{fig:lep_mssm}The 95 \% C.L. exclusion contours 
         in the $\mathrm{(\tan\beta,m_h)}$ projection for 
         the ``$\mathrm{m_h-max}$'' scenario.  
         The results of the four LEP experiments are combined.
         The hatched area is excluded, the filled area 
         is theoretically inaccessible and the dashed line
         indicates the expected boundary of the excluded region
         in the absence of a signal.
         }
\end{minipage}
\end{figure}

At LEP, searches for the neutral Higgs bosons of the 
MSSM are performed, exploiting two production mechanisms:
the Higgs-strahlung and associated Higgs boson pair production,
$\mathrm{e^+e^-\ra Z^*\ra hA}$. 
For the Higgs-strahlung process, the
same event topologies as in the SM Higgs search are studied.
For the hA production, the final states studied are:
$\mathrm{hA\ra b\bar{b}b\bar{b}}$ and 
$\mathrm{hA\ra b\bar{b}\tau^+\tau^-,\tau^+\tau^- b\bar{b}}$.
To improve search sensitivity in the MSSM parameter regions 
where the decay $\mathrm{h\ra AA}$ opens, some collaborations
have searched also for the $\mathrm{e^+e^-\ra hZ\ra AAZ}$
process~\cite{lo_mssm}. 
No evidence for the production of the neutral MSSM Higgs bosons 
is found in the LEP combined data~\cite{mssm_lep}. 
Hence, the search results are
interpreted in terms of exclusion of the MSSM parameter regions.
Figure~\ref{fig:lep_mssm} shows the 95\% C.L. excluded regions in the 
$\mathrm{(\tan\beta,m_h)}$ plane for the so-called ``$\mathrm{m_h-max}$''
scenario, which yields the maximal theoretical upper limit 
on $\mathrm{m_h}$ in the model.
In this scenario, the LEP combined data establish the 
95\% C.L. lower limits on the h and A boson masses of 
91.0 GeV and 91.9 GeV, respectively.

\begin{figure}[h]
\begin{center}
\includegraphics[width=0.85\textwidth]{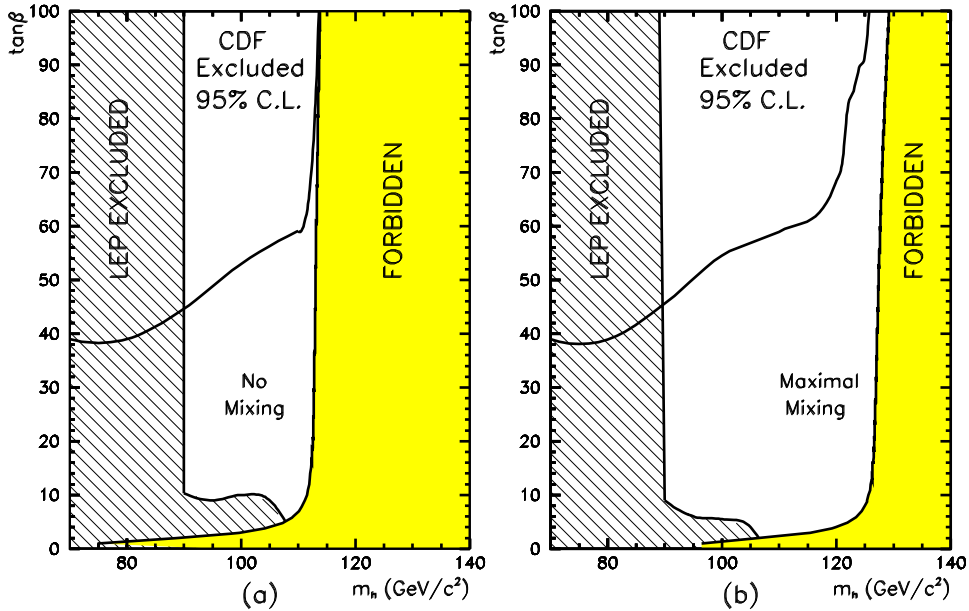}
\end{center}
\caption{\label{fig:mssm_cdf}
The 95\% C.L. excluded regions in the $\mathrm{(\tan\beta,m_h)}$ plane
for the two scalar top mixing scenarios : a) ``no mixing'', 
b) ``maximal mixing''. Results are obtained by the CDF collaboration.
The LEP excluded regions shown here are obsolete.}
\end{figure}

Using RunI data, the CDF collaboration has 
carried out the search for the neutral 
supersymmetric Higgs bosons in the process 
$\mathrm{p\bar{p}\ra b\bar{b}\phi\ra b\bar{b}b\bar{b}}$ 
with $\mathrm{\phi=h,H}$ and A, exploiting the enhanced 
Higgs-bottom Yukawa coupling at high values of $\tan\beta$~\cite{mssm_cdf}. 
A dedicated analysis is elaborated to select events with four or more 
jets of which at least three being tagged as b jets. 
The study of dijet invariant mass spectra in the selected sample revealed
no evidence for a signal. Shown in 
Figure~\ref{fig:mssm_cdf} are the exclusion contours in the 
$\mathrm{(\tan\beta,m_h)}$ plane for the two 
representative scenarios of the scalar top mixing.

\section{Conclusions}
Searches for Higgs bosons of various theoretical 
models are performed at LEP and Tevatron.
No strong indication of the production of Higgs bosons 
is found. Negative results of the searches are used to 
constrain parameters in the Higgs sector of the SM and 
its various extensions. In the SM, the 95\% C.L. 
lower limit of 114.4 GeV on the Higgs boson mass is 
obtained from the direct searches at LEP.

Searches for Higgs bosons will be one of the main objectives
during the RunII data taking period at Tevatron. 
Higgs searches will be also continued 
at the Large Hadron Collider (LHC) at CERN. If the Higgs
boson is found, its profile will be explored at a 
future linear $\mathrm{e^+e^-}$ collider.
The physics results from these 
experimental facilities are anticipated to shed a light on 
the problem of electroweak symmetry breaking and give an
insight into the origin of mass.

\section*{Acknowledgements}
I thank Dr. Wolfgang Lohmann for giving talk 
at Physics in Collision Conference on behalf of my 
name and for his help in preparation of this paper.

\end{document}